\documentclass[prb,aps,preprint,showpacs]{revtex4}
\usepackage{revsymb,amsmath}
\usepackage{graphicx}
\usepackage{amsmath,bm}
\begin{document}
\author{Michael Schulz }
\affiliation{Universit\"at Ulm\\D-89069 Ulm Germany}
\email{michael.schulz@uni-ulm.de}
\author{Steffen Trimper}
\affiliation{Institute of Physics,
Martin-Luther-University, D-06099 Halle Germany}
\email{steffen.trimper@physik.uni-halle.de}
\title{Thermally Assisted Spin Hall Effect}
\date{\today }

\begin{abstract}
The spin polarized charge transport is systematically analyzed as 
a thermally driven stochastic process. 
The approach is based on Kramers' equation describing the semiclassical motion under the inclusion of stochastic 
and damping forces. Due to the relativistic spin-orbit coupling the damping experiences a relativistic correction 
leading to an additional contribution within the spin Hall conductivity. A further contribution to the 
conductivity is originated from the averaged underlying crystal potential, the mean value of which depends 
significantly on the electric field. We derive an exact expression for the electrical conductivity. All corrections 
are estimated in lowest order of a relativistic approach and in the linear response regime.
\end{abstract}


\maketitle

\section{Introduction} The scattering of an unpolarized electron beam at a likewise unpolarized target leads to a separation 
of electrons with different spins \cite{m}. The reason for that effect is the spin-orbit interaction originated by 
relativistic corrections to the Schr\"odinger equation of the order $c^{-2}$. As a consequence an electric 
current in a semiconductor should be accompanied by a spin current perpendicular to the conventional current 
of the charge carriers \cite{dp}. The generation of a transverse spin current due to an external 
electric field called as spin Hall effect \cite{h} has attracted much attention recently particularly for realizing 
spintronic devises \cite{w}. In a series of experiments \cite{k,s,wu,stern} the effect had been observed in 
different materials. Recently the spin Hall conductivity was measured at room temperature \cite{stern}, 
an observation which give the motivation for the present paper. The transport process is considered as a 
thermally activated process which is additionally driven by the applied electric field. This field however 
reveals a direct coupling to the crystal potential in such a manner that also in case of a periodic potential 
a constant current is maintained. Our approach is based on the Fokker-Planck equation or Kramers' equation 
for the semiclassical motion of 
the charge carriers with damping and coupling to a stochastic source like a heat bath. Thus the charged particles 
can overcome permanently, but with a certain probability, potential barriers. Even for low temperature one should 
take into account the coupling of the homogeneous electric field to the potential landscape. Actually as discussed 
below the averaged force acting on the particles consists of several parts which point at different directions and 
contribute to the total conductivity. Recently a very promising theory has been proposed \cite{c} by studying 
the semi-classical deterministic equation of motion. The theoretical approach describing the spin Halle effect 
goes back to \cite{dp} where a mesoscopic equation for the spin density had been proposed. A more phenomenological 
approach based on an anomalous scattering mechanism in the absence of spin-flip scattering discussed in \cite{h}. 
As a consequence anomalous currents and a finite spin-diffusion length had been studied in \cite{z} where 
the calculations are based on the Boltzmann equation. In a series of papers \cite{mnz,erh,hv,ycz,lcc} a quantum 
approach for the spin Hall conductivity had been discussed. In \cite{psm} a classical theory is proposed.\\ 
In the present paper we consider an extension of the theoretical approach proposed in \cite{c,psm} by including 
systematically a coupling to stochastic forces and friction terms. Because we are in the linear response regime 
the canonical friction and the strength of the stochastic force are related by the fluctuation-dissipation theorem. 
The steady state solution of Kramers' equation is an appropriate tool to obtain the electrical current. 
As a novelty we discuss in detail the influence of the underlying crystal potential. Unlike \cite{c} we get 
corrections to the spin Hall conductivity by the averaged crystal potential. Because the mean value over the crystal 
potential is performed with the stationary solution of Kramers' equation and this solution depends on the 
electric field there occur additional corrections to the conductivity in both the linear response regime 
with respect to the field $\bf E$ and in lowest order $c^{-2}$ of an relativistic approach. Whereas the 
relation for the conductivity under inclusion of the crystal potential is an exact result, the real contribution,  
originated from the potentials, has to be estimated by model potentials.\\

\section{Model} 
In lowest order of relativistic effects the Hamiltonian has to be supplemented by the 
spin-orbit interaction 
\begin{equation}
\mathcal{H} = \frac{\bf p^2}{2m} + U(\bf r) + \kappa\, {\bf p} \cdot \left[ {\bm \sigma} \times \nabla U \right]
\quad\rm{with}\quad \kappa = \frac{{\hbar}}{4m^2 c^2}
\label{ham}
\end{equation}
The quantity $\bm \sigma$ are Pauli operators which are assumed to be constant. Due to the spin-orbit 
interaction the total potential $U(\bf r)$, related to all forces acting on the charged particles, 
appears twofold. In the present paper we incorporate in the potential the intrinsic crystal potential 
$U_c({\bf r})$ and the energy of the homogeneous electric field  $U({\bf r}) = U_c({\bf r}) - e {\bf E} \cdot {\bf r}$. 
Extending the deterministic approach proposed in \cite{c} let us include stochastic forces ${\bm \eta}({\bf r},t)$ 
due to other degrees of freedom as imperfections. Additionally, the charged particles are subjected to an arbitrary 
damping force  ${\bm \Gamma}({\bf r},{\bf p})$. Under that conditions the semiclassical equations of motion read
\begin{eqnarray}
\dot{\bf r} &=& \frac{{\bf p}}{m} + \kappa {\bm \sigma} \times \nabla U;\nonumber\\
\dot{\bf p} &=& -\nabla U - \kappa \nabla 
\left[ {\bf p}\cdot ({\bm \sigma} \times \nabla U)\right] - \Gamma({\bf r},{\bf p}) + {\bm \eta} (t)\,.
\label{kan}
\end{eqnarray} 
Because we are interested in the linear response regime the system is assumed to be nearby the equilibrium. Therefore 
it seems to be adequate to assume the stochastic force $\bm \eta$ distributed by a Gaussian white noise with an 
arbitrary noise strength $D$: 
$\langle \eta_{\alpha}(t)\,\eta_{\beta}(t') \rangle = 2\,D_{\alpha \beta} \delta(\,t - t')\,$. 
The corresponding Fokker-Planck or Kramers' equation for the probability density $P({\bf r}, {\bf p}, t)$ 
is \cite{gar}
\begin{equation}
\frac{\partial P}{\partial t} = - \frac{\partial }{\partial {\bf r}}
\left[ \frac{\partial \mathcal{H}}{\partial {\bf p}} P  \right] + 
\frac{\partial }{\partial {\bf p}}
\left[ \left(\frac{\partial \mathcal{H}}{\partial {\bf r}} + {\bm \Gamma}
\right) P  \right] 
+ D \frac{\partial ^2}{\partial {\bf p}^2} P
\label{fp}
\end{equation}
It is easy shown that the last equation satisfies the principle of detailed balance  
and exhibits an equilibrium solution of the form 
\begin{equation}
P_e({\bf r}, {\bf p}) \sim \exp (-\frac{\mathcal{H}}{k_B T} )\,,
\label{eq}
\end{equation}
provided the arbitrary damping function ${\bm \Gamma}$ fulfills the relation
\begin{equation}
{\bm \Gamma} = \frac{D}{k_BT} \frac{\partial \mathcal{H}}{\partial {\bf p}}\,.
\label{fric}
\end{equation}
From here we conclude immediately  
\begin{equation} 
{\bm \Gamma} = \gamma\, {\bm \Pi },\quad {\rm with}\quad {\bm \Pi } =  \left[ {\bf p} +  m \kappa (\,{\bm \sigma} \times \nabla U(\bf r)) \right]
\label{fric1}
\end{equation}
The quantity ${\bm \Pi}$ is the canonical momentum whereas the parameter $\gamma$ is the damping constant or 
the inverse relaxation time, which appears in the conventional Einstein relation $D = m \gamma k_B T\,$. 
Due to the linear momentum part in the Hamiltonian \eqref{ham} the 'canonical' friction includes likewise a 
relativistic correction, which leads to an additional term in the Drude conductivity discussed below.\\ 

\section{Stationary current}
Owing to the electric field energy 
incorporated into the Hamiltonian Eq.~\eqref{ham}, the equilibrium distribution Eq.~\eqref{eq} may not be  
appropriate for analyzing the electric current. All charged particles would shift to $-\infty$ to establish 
an equilibrium state. Such a state is not realized because the equilibrium distribution is not normalizable 
in that case. Instead of the equilibrium solution one needs a steady state  
solution. Due to the conservation of the probability Eq.~\eqref{fp} can be written in form 
of a continuity equation where the probability current is a function of both, the spatial coordinate and the 
momentum ${\bf j}_p({\bf p},{\bf r},t)$. Whereas the equilibrium distribution $P_e$ satisfies ${\bf j}_p = 0$ a steady 
state solution $P_s$ obeys the weaker condition $\nabla \cdot {\bf j}_p = 0$. The complete evolution equation for the 
problem reads now  
\begin{eqnarray}
\frac{\partial P}{\partial t} &=& - \frac{\partial }{\partial {\bf r}} \left( {\bf A} P \right )
- \frac{\partial }{\partial {\bf p}} \left( {\bf B} P \right )
+ D \frac{\partial ^2}{\partial {\bf p}^2} P \quad{\rm with} \nonumber\\
{\bf A} &=& \frac{{\bf p}}{m} + \kappa {{\bm \sigma}} \times \nabla U\nonumber\\
{\bf B} &=& - \nabla U - \kappa \nabla \left[ {\bf p} \cdot ({\bm \sigma} \times \nabla U )\right] - 
{\bm \Gamma}({\bf r}, {\bf p})  
\label{fp1}
\end{eqnarray}
From here we conclude
\begin{equation}  
\frac{\partial}{\partial t}\langle {\bm \Pi} \rangle = \langle {\bf B} \rangle + 
m \kappa \langle {\bf A} \cdot \nabla ({\bm \sigma} \times \nabla U({\bf r}) \rangle\,,\quad 
\frac{\partial}{\partial t}\langle {\bf r} \rangle = \langle {\bf A} \rangle\,.
\label{stat}
\end{equation}
To maintain a stationary electric current $\bf j$ the averaged velocity of the charge 
carriers should be fixed. In according to Eq.~\eqref{stat} this requirement is fulfilled by   
\begin{equation}
{\bf j} = n e \langle\, {\bf A }\,\rangle\,,\quad \frac{\partial}{\partial t} \langle {\bm \Pi} \rangle = 0\,.
\label{cur}
\end{equation}
The quantity $n$ is the density of the charge carriers. The bracket means now the average with the steady 
state solution $P_s({\bf r, p};\,{\bf E},\kappa)$ of Eq.~\eqref{fp1}. It is important to note that all 
averages performed with $P_s$ depends on the electric field $\bf E$, too. Because we are looking for a 
solution linear in the relativistic factor $\kappa$ we consider firstly the non-relativistic case $\kappa \equiv 0$. 
From Eq.~\eqref{stat} combined with Eq.~\eqref{cur} we get
\begin{equation}
\gamma\, \langle {\bf p} \rangle = e {\bf E} - \langle \nabla U_c \rangle\,.  
\label{av}
\end{equation}
Since in the non-relativistic case, $\kappa = 0$ the system offers only one preferential direction, 
namely that one given by the external field $\bf E$, and furthermore the distribution function of 
the steady state depends on the electric field, the only possibility is that the averaged force  
in Eq.~\eqref{av} point at the field direction. Hence we set 
$$
\langle \nabla U_c \rangle = \alpha_1 \,e\, {\bf E} \,,
$$
where the dimensionless factor $\alpha_1$ characterizes the influence of the crystal potential. 
The factor $\alpha_1$ depends on the concrete form of the potential and will be discussed below 
for a Coulomb potential. Inserting the last relation in Eq.~\eqref{cur} it results in the non-relativistic case
\begin{equation}
{\bf j}_0 = \frac{e^2 n}{m \gamma} (1 - \alpha_1)\, {\bf E} \,.
\label{dru}
\end{equation}
The first part is nothing else as the conventional Drude conductivity with the simplest form of a scattering mechanism 
included in the damping factor $\gamma$. The charge carriers will follow that law if the influence of the 
crystal potential is negligible, a situation which should be realized at sufficient high temperatures. 
In case the lattice potential becomes relevant an additional scattering mechanism is established and the 
mobility of a charge particle is further reduced leading to a decrease of the conductivity. 
Now let us take into account the relativistic effects in lowest order in $\kappa$.  From Eqs.~\eqref{stat},\eqref{cur} 
\begin{equation}
{\bf j} = \frac{n e^2}{m \gamma} \left [ {\bf E} - \frac{1}{e} \langle \nabla U_c \rangle \right] 
+ \frac{2 \kappa n e^2 \,C\,(1 - \alpha_1\,)}{m \gamma^2}  ({\bm \sigma} \times {\bf E})\,, 
\label{cur1}
\end{equation}
where the parameter $C$ is defined by 
\begin{equation}
\left \langle \frac{\partial^2 U_c}{\partial x_{\mu } \partial x_{\nu }} \right \rangle
= C\, \delta_{\mu \nu }\,.  
\label{A}
\end{equation}
Notice that the factor $C$ differs by a factor $e$ from that defined in \cite{c}. The last relation is obviously 
valid for an isotropic crystal. As observed in Eq.~\eqref{cur1} the relativistic effect proportional to $\kappa$ 
yields contributions to the current $\bf j$ pointing toward the field direction $\bf E$ and perpendicular to that 
one in the ${\bm \sigma} \times {\bf E}$ direction. Hence there exists a third direction oriented toward  
${\bm \sigma} \times ( {\bm \sigma} \times {\bf E})$. Because the average over the crystal potential 
has to be fulfilled with the stationary solution of the Kramers equation and this distribution depends on 
the electric field, the most general expression for the averaged force linear in the field $\bf E$ reads 
\begin{equation}
\left \langle \nabla U_c \right \rangle = \alpha_1\, e {\bf E} + \kappa e 
\left[\,\alpha_2\,( {\bm \sigma} \times {\bf E}\,) + 
\,\alpha_3\,( {\bm \sigma} \times ( {\bm \sigma} \times {\bf E})\,)\,\, \right]\,.
\label{cur2}
\end{equation}
Here the parameters $\alpha_i,\,i\,= 1,\,2\,,3$ reflect the influence of the crystal potential. They 
will be estimated later. As the final result the electric current consists of three parts
\begin{eqnarray}
{\bf j} &=& \frac{n e^2}{m \gamma} (1 - \alpha_1) {\bf E} + 
\frac{n e^2 \kappa \alpha_3}{m \gamma} \left[ ({\bm \sigma} \times {\bf E}) \times {\bm \sigma} \right]\nonumber\\
&+& \frac{ n e^2 \kappa}{m \gamma^2} 
\left[ 2 \,C\,(1-\alpha_1) - \alpha_2 \gamma \right ] ({\bm \sigma} \times {\bf E})\,.
 \label{cur3}
\end{eqnarray}
In the subsequent section we discuss the conductivity.\\

\section{Spin Hall conductivity} 
\begin{figure}[h]
\centering
%\psfrag{p}[][][2.1]{$\delta p$}
\includegraphics[angle=-90,width=\linewidth]{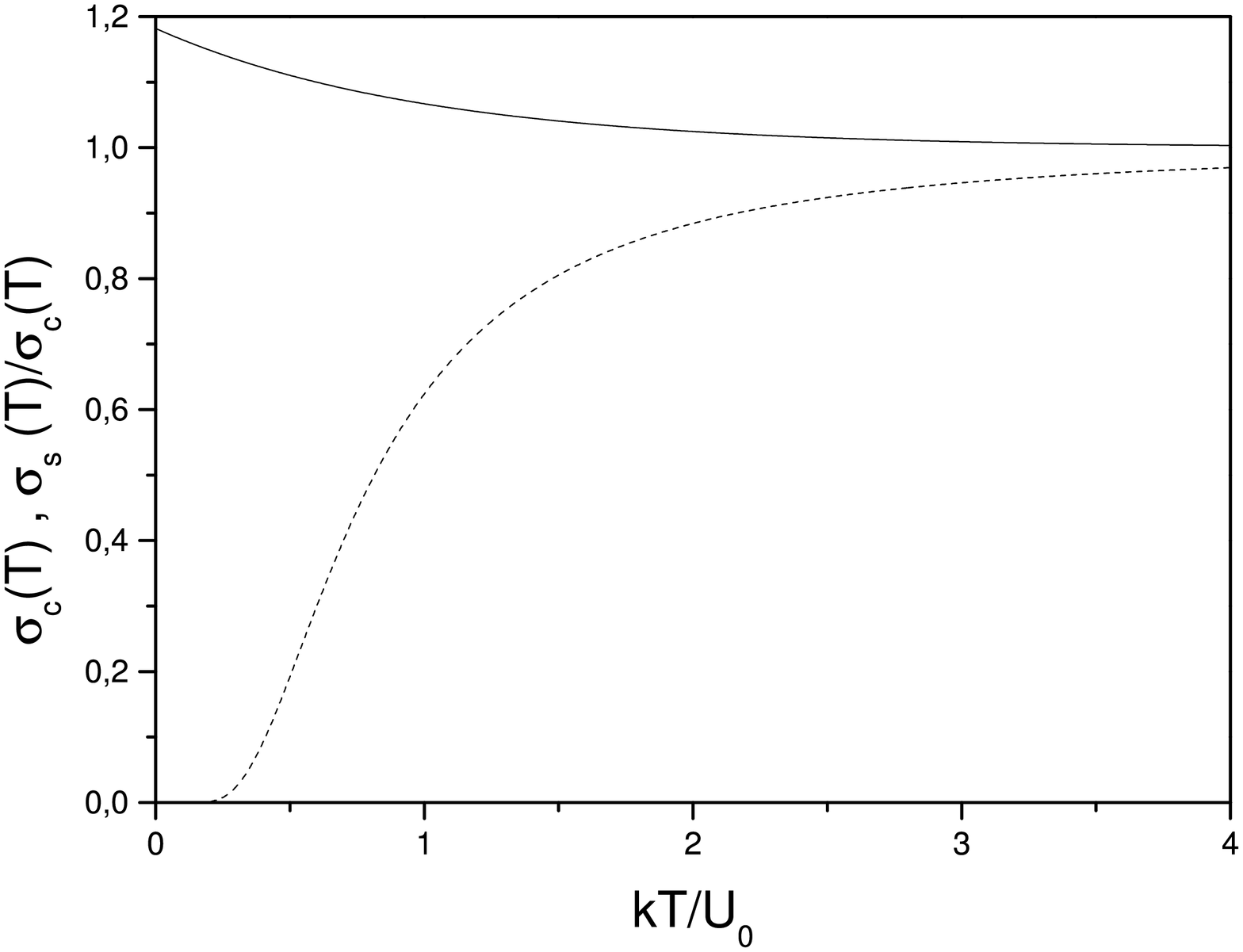}
\caption{The charge conductivity in units of the classical Drude conductivity versus the rescaled temperature 
(dashed line) and the ratio $\sigma_s/\sigma_c$ in units of $\hbar C/4m^2c^2\gamma$ (full line)}
\label{fig10}
\end{figure}

From Eq.~\eqref{cur3} one obtains the charge conductivity and the Hall conductivity 
\begin{equation}
\sigma_c = \frac{n e^2 }{m \gamma} (1 - \alpha_1)\,,\quad \sigma_s = \frac{ n e^2 \kappa}{m \gamma^2}\, 
\left[ 2\,C (1 - \alpha_1) - \alpha_2\,\gamma \right]\,.
\label{cond}
\end{equation}
The parameter $C$ is defined in Eq.~\eqref{A} whereas the coefficients $\alpha_i$ are introduced in Eq.~\eqref{cur2}. 
As already observed in \cite{c} the ratio
\begin{equation}
\frac{\sigma_s}{\sigma_c} = \frac{\hbar}{4 m^2 c^2} \frac{\, 2\, C\,(1 - \alpha_1 ) - \alpha_2\,\gamma }{(1 - \alpha_1)\gamma }
\label{ratio}
\end{equation}
is independent on the concentration of the charge carriers and is determined by the periodic potential. 
The result is in accordance with microscopic models \cite{erh}. The expression for the electrical current 
is different from that obtained previously \cite{c}. However, if all parameters $\alpha_i$ are neglected the ratio 
in Eq.~\eqref{ratio} is in accordance with \cite{c}. Apart from the thermal activation process as driving force 
the conventional Drude current is already supplemented by a correction term due to the crystal potential, 
see ${\bf j}_0$ in Eq.~\eqref{dru}. Otherwise the spin Hall current is likewise corrected twice by the relativistic  
damping force $\bm \Gamma$ and through the influence of the lattice potential. To discuss the differences in more 
detail let us estimate the correction coefficients $C,\,\alpha_i\,$. The parameter $C$ is originated by the 
distribution of charge carriers in a cubic lattice. Following the line proposed in \cite{c} one can estimate the 
coefficient $C$ using the Poisson equation 
$$ 
\nabla^2 U_c(\bf r) = - 4 \pi \rho e
$$
where $\rho$ is the local charge density of the ions. This gives 
\begin{equation}
C = \frac{4\pi}{3}Z e n_0
\label{cor1}
\end{equation}
Here $-Z e$ is the charge and $n_0$ is the concentration of the ions. Let us notice that the result is correct in 
linear order of the external field. Using the approach proposed below one can show that corrections  
to the parameter $C$ due to the field $\bf E$ occur at first in second order. The remaining quantities denoted by 
$\alpha_i$, may be figured out in a more laborious analysis. In finding out the parameters we have to 
calculate $\langle \nabla U_c \rangle$ according to Eq.~\eqref{cur2}. 
The problem one is confronted with is the fact that the electric field affects the intrinsic crystal potential 
$U_c({\bf r})$ in such a manner that a constant current is maintained. In particular, the coupling between the external 
field and the crystal potential becomes important for low excited charge carriers which are subjected strongly 
to the field.  In our approach the influence of the homogeneous field $\bf E$ on the potential is manifested 
by the explicit field dependence of the steady state solution $P_s({\bf r}, {\bf p};{\bf E}, \kappa)$. As a 
direct consequence the averaged force carried on the charge particles via the crystal potential, 
depends on $\bf E$ according to Eq.~\eqref{cur2}. Owing to the pure spatial dependence of the crystal 
potential the reduced distribution function, defined by 
$\rho({\bf r}; {\bf E}, \kappa) = \int d{\bf p} P_s({\bf r}, {\bf p};{\bf E}, \kappa)\,$
seems to be a more appropriate quantity as $P_s$. However the basis Kramers' equation \eqref{fp1} 
allows no exact steady state solution $P_s$. Hence one is relied on a reasonable approximation for 
the stationary distribution function. We are looking for a solution of Eq.~\eqref{fp1} 
in the linear response regime of the external field and in lowest order of the relativistic factor $\kappa$. 
Let us make the ansatz for the steady state solution of Eq.~\eqref{fp1} in the form 
\begin{equation}
P_s({\bf r}, {\bf p}; {\bf E}, \kappa) = P_0({\bf r}, {\bf p}; {\bf E}) 
\left[ 1 + \kappa {\bf p}\cdot {\bf f}({\bf r, E}\,) \right]\,,
\label{app1}
\end{equation}
where ${\bf f} = - \frac{1}{k_B T} {\bm \sigma} \times \nabla U_c\,$ in lowest order. 
Inserting this ansatz in Eq.~\eqref{fp1}, 
performing the momentum integration and collecting all terms in linear order in $\bf E$ and 
$\kappa$ one obtains 
$$
\langle \nabla U_c \rangle = \int d{\bf r} \left [ \rho ({\bf r, E}, \kappa = 0) +  
\kappa \overline{\bf p({\bf r, E})} \cdot f ({\bf r, E}) \right ] \nabla U_c \,.
$$
Here the quantity  $\overline{\bf p({\bf r, E})} = \int d{\bf p} {\bf p} P_0({\bf r}, {\bf p}; {\bf E})$
is the averaged momentum proportional to the current at the coordinate $\bf r$ for $\kappa = 0$. 
Assuming that the reduced distribution function $\rho$ obeys the Smoluchowski equation \cite{gar} 
and making the approximation 
\begin{equation}
\left \langle \frac{\partial U_c}{\partial x_{\alpha}} \frac{\partial^2 U_c}{\partial x_{\beta}\partial x_{\gamma}} \right \rangle 
\simeq  C\, \delta_{\beta \gamma}\,\left \langle \frac{\partial U_c}{\partial x_{\alpha}}\right \rangle 
\label{cc}
\end{equation}
we find
$$
\langle \nabla U_c \rangle = \alpha_1 e {\bf E} + \frac{ e \kappa C ( 1- \alpha_1)}{\gamma}  ({\bm \sigma} \times {\bf E}) \,.
$$
Comparing the result with the ansatz made in Eq.~\eqref{cur2} we conclude
\begin{equation}
\alpha_2 = \frac{ ( 1- \alpha_1 ) C}{\gamma}, \quad \alpha_3 = 0
\label{al22}
\end{equation}
In lowest oder of a relativistic approach there is no current in spin direction. Inserting this result in 
Eq.~\eqref{ratio} one obtains that the ratio of the spin Hall conductivity and the charge conductivity is 
independent on the lattice potential
\begin{equation}
\frac{\sigma_s}{\sigma_c} = \frac{\hbar C}{4 m^2 c^2 \gamma}\,. 
\label{fin}
\end{equation}
This expression differs from that obtained in \cite{c} by a factor 2. The reason for the discrepancy is 
that in \cite{c} the field dependence of the averaged crystal potential is neglected. The spin and the charge 
current is subjected to the same scattering mechanism by the 
underlying isotropic crystal potential. If one includes higher order corrections 
appearing in Eq.~\eqref{cc} then the ratio should be dependent on the potential. Making a similar approach 
one can derive the remaining 
factor $\alpha_1$. 
We find in $d$ dimensions
\begin{equation}
\alpha_1 = \frac{1}{ V\,d\, (k_B T)^2} \int d{\bf r} \left( U_c({\bf r}) - \overline{U_c} \right )^2,\quad
\overline{U_c} = \frac{1}{V} \int d{\bf r}\, U_c({\bf r}) 
\label{cor2}  
\end{equation}
Notice that the result is valid in first order of $\kappa$ and in the linear response regime. 
So the problem in finding out the influence of the crystal potential is reduced to calculate 
the mean square displacement according to Eq.~\eqref{cor2}. To that aim we have considered a  
Coulomb potential $U_c = Z e^2 \sum_{i,j} { \mid {\bf r}_i  - {\bf r}_j \mid }^{-1} $. 
A straightforward calculation yields
$$
\alpha_1 = \frac{4 (Z e^2)^2 n a}{3 (k_B T)^2 \pi}\, \equiv  \left(\frac{2 \epsilon_A}{\sqrt{3 \pi} k_B T} \right)^2 
$$
where we have assumed a simple cubic lattice with the lattice spacing a. 
There occurs a characteristic energy $\epsilon _A =  Z e^2 \sqrt{n a}$. 
For high temperatures $\epsilon_A \ll k_B T$ the crystal potential becomes irrelevant, i. e the motion of 
the charge carriers is independent on the 
underlying potential. Otherwise our approach is a classical one which should be restricted to 
$\epsilon_A \leq k_B T$. If the density $n$ of the charge carriers is of the order $n \simeq a^{-3}$ then 
$\epsilon_ A \simeq  \frac{Z e^2}{a}$. For a typical electronic system the thermal energy is of 
the order of the Fermi energy $k_B T \simeq \epsilon_F$ and therefore 
$$
\sigma_c \simeq \frac{n e^2}{m \gamma} \left[ 1 - \left(\frac{2 \epsilon_A}{\,\sqrt{3 \pi}\,\epsilon_F} \right)^2 \right]\,. 
$$
Even in case that $\epsilon_A \simeq \epsilon_F$ there is a significant contribution due to the potential. In general 
the characteristic energy should be smaller than the Fermi energy.  
Instead of using a Coulomb potential one can also consider a $\cos$-potential $
U_c({\bf r}) = U_0 \sum_{i=1}^{d} (1 - \cos (q_i x_i) )\,$. 
For this potential one can calculate the conductivity numerically. 
The ratio of the charge and the spin Hall conductivity versus the rescaled temperature are shown in Fig.~\ref{fig10}. 
For high temperature the potential becomes irrelevant and the classical Drude conductivity is achieved, 
compare the dashed line in Fig.~\ref{fig10}. The reason for the weak deviations of $\sigma_s/\sigma_c$ 
from Eq.~\eqref{fin} is originated exclusively by the approximation made in Eq.~\eqref{cc}.\\ 

\section{Conclusion} 
We have presented a thermally driven Drude model under inclusion of the relativistic 
spin-orbit coupling. The charge carriers are driven by a homogeneous electric field and a Gaussian white noise. 
Additionally the charged particles are subjected to an intrinsic periodic crystal potential. Our approach 
differs from the semiclassical description proposed in \cite{c} in two respects. Firstly, we start consequently 
from the stochastic Kramers' equation including canonical friction and white noise terms, and secondly we take 
into account that the correction terms due to the averaged crystal potential exhibit a pronounced dependence 
on the the electric field and its orientation with respect to the spins. Our approach shows that Spin Hall Effect 
should also observed for finite temperature as it has reported recently \cite{stern}. For high temperatures 
the particles are not influenced by the potential and the conventional Drude conductivity appears. When the 
temperature is lowered the mobility is reduced owing to the potential. In our approach the interplay between 
the electric field and the lattice potential is manifested in the field dependence of the averaged force imposed on 
charged particles. \\ 

\noindent The work has been supported by the DFG: SFB 418

\clearpage

\end{document}